\documentclass[aps,prl,twocolumn,groupedaddress]{revtex4}

\usepackage[dvips]{graphicx}
\usepackage[]{amsmath}
\usepackage[]{amssymb}
\usepackage[]{amsfonts}
\usepackage[]{latexsym}

\newcommand{\mb}[1]{\mbox{\boldmath $#1$}}

\newcommand{\eas}[0]{\begin{eqnarray*}}
\newcommand{\eae}[0]{\end{eqnarray*}}
\newcommand{\les}[0]{\begin{equation}}
\newcommand{\lee}[0]{\end{equation}}
\newcommand{\leas}[0]{\begin{eqnarray}}
\newcommand{\leae}[0]{\end{eqnarray}}
\newcommand{\mat}[4]
{
\left(
\begin{array}{cc}
#1 & #2 \\
#3 & #4 
\end{array}
\right)
}
\newcommand{\mvec}[2]
{
\left(
\begin{array}{c}
#1  \\
#2  
\end{array}
\right)
}
\begin{document}


\title{Superconductivity and Abelian Chiral Anomalies} 


\author{Y. Hatsugai}
\email[]{hatsugai@pothos.t.u-tokyo.ac.jp}
\author{S. Ryu}
\affiliation{Department of Applied Physics, University of Tokyo,\\
7-3-1, Hongo, Bunkyo-ku, Tokyo 113-8656, JAPAN}
\author{M. Kohmoto}
\affiliation{
Institute for Solid State Physics, 
University of Tokyo, 
5-1-5, Kashiwanoha, Kashiwa, Chiba, 277-8581,JAPAN}


\date{Nov. 11, 2003}

\begin{abstract}
Motivated by the geometric character of spin Hall conductance,
the topological invariants of generic superconductivity
are discussed  based on the 
Bogoliuvov-de Gennes equation on lattices.
 They are given by the Chern numbers of degenerate
condensate bands for unitary order, which are 
realizations of Abelian chiral anomalies for non-Abelian connections.
The three types of Chern numbers for the $x,y$ and $z$-directions are 
given by covering degrees of some doubled surfaces
 around the Dirac monopoles.
For nonunitary states,
several topological invariants
 are defined by analyzing the so-called  $q$-helicity.
Topological origins of the
nodal structures of superconducting gaps are also discussed.
\end{abstract}


\maketitle

%



The importance of quantum-mechanical phases
in condensed matter physics has been recognized and emphasized 
for recent several decades.
The fundamental character of a
vector potential 
is evident in the Aharonov-Bohm effect
where  the $U(1)$ gauge structure is essential 
and  a magnetic field in itself plays only  a secondary role
\cite{ab}.
Topological structures in 
quantum gauge field theories have also been 
studied and extensive knowledge 
has been  accumulated\cite{eguchi}.
Quantum mechanics itself 
supplies a fundamental gauge structure\cite{berry}.
It is known as  geometrical phases in many different contexts,
where gauge structures emerge by restricting  physical 
spaces.
The quantum Hall effect is one of 
the key phenomena to establish 
the importance  of geometrical phases\cite{tknn}.
The topological 
character of the Hall conductance was first
realized  by the Chern number expression,
where the Bloch functions define ``vector potentials'' in 
the magnetic Brillouin zone 
accompanied with a  novel gauge structure\cite{mkann}.
Further the ground state of the
fractional quantum Hall effect is a complex many-body state
where another kind of  gauge structure emerges\cite{fqh_l}.
These quantum states with nontrivial 
geometrical phases are
 characterized by  topological orders
which 
extend an idea of order parameters in
statistical mechanics to 
the quantum states  without
spontaneous symmetry breaking
\cite{wen_topological1}.
Also we point out an importance of boundary effects for 
topologically nontrivial systems. 
Bulk properties are closely
related to edge states and 
localized states near impurities
and vortices\cite{edge_l,edge_yh,chern_yh,yh-qhe-review,sr-yh}.

Topologically nontrivial structures in superconductors
also have a long history.
Recently, following a prediction of flux phases for correlated 
electron systems\cite{am},
spin Hall conductance is defined for 
superconductors based on the Bogoliuvov-de Gennes (B-dG) 
equation\cite{smf,ym-yh-pairing,tesa}
\footnote{ 
A nontrivial gauge structure in superfluid $^3{\rm He}$
was discussed from a different point of view\cite{vl-y,vol-book}.
}.
As for singlet states and triplet states besides equal-spin-pairing
states,
a map to a parameter space which represents the BdG hamiltonian
is considered.\cite{yh-sr}
In the parameter space, an analogue of
 the Dirac monopole \cite{dirac} exists and 
the Chern numbers are analyzed.

In this paper, we establish a topological 
characterization of 
{\em general } superconductivity
  based on the 
BdG equation on  lattices.
The energy spectrum of the BdG hamiltonian are
fully used to calculate the Chern numbers
of the superconductors.
As for the unitary superconductors,
condensed matter realizations of 
chiral anomalies for non-Abelian connections
are given explicitly.
Topological consideration is useful to 
distinguish
superconductivities with the same pairing symmetry.
The present analysis also clarifies 
nodal  structures of superconducting gaps
with various anisotropic order parameters,
which is closely related to the quantum Hall effect
in three-dimensions\cite{h3d,3dq}.
Various types of the nodal structures are not accidental
but have fundamental topological origins.
A possible time-reversal symmetry-breaking 
and an unconventional gap structure are proposed based on the 
experiments
\footnote{ 
Possible point nodes in superconductivities of 
the filled-skutterudite PrOs$_4$Sb$_{12} $ associate with
a breaking of a time-reversal symmetry are discussed in
\cite{izawa,aoki}.
}

{\it Bogoliuvov-de Gennes hamiltonian}.
Let us start from the following
 hamiltonian on lattices with spin-rotation symmetry:
\begin{alignat*}{1} 
H =
&
\sum_ {ij } t_{ij} c_{i \sigma  } ^\dagger c_{j \sigma  }
+
\sum_{ij}
V_{ij}^{\sigma_1 \sigma_2;\sigma _3 \sigma _4 } 
c ^\dagger _{i \sigma_1  } c ^\dagger _{ j \sigma _2 }
c_{j \sigma _3  } c_{i \sigma _4 } -\mu\sum_{i  }c_{i \sigma } ^\dagger c_{i \sigma }
\end{alignat*} 
where $c_{i\sigma}$ is the electron annihilation operator with  spin $\sigma$ at site 
$i$,
$
t_{ij} = t_{ji}^*$,
$V_{ij}^{\sigma_1 \sigma_2;\sigma _3 \sigma _4 } 
= 
{(V_{ij}^{ \sigma_4 \sigma_3;\sigma _2 \sigma _1 } )}^*
$, 
$ 
V_{ij}^{\sigma_1 \sigma_2;\sigma _3 \sigma _4 } 
= 
V_{ji}^{\sigma_2 \sigma_1;\sigma _4 \sigma _3 } 
$ and $\mu$ is a chemical potential. 
Summations over repeated spin indices $\sigma$ are implied hereafter.
The mean field BCS approximation leads to,
\begin{alignat*}{1} 
 {\cal H} 
=& 
 \sum_ {ij } t_{ij} c_{i \sigma  } ^\dagger c_{j \sigma  }
+
\sum_{ij  }
(
\Delta _{ij} ^{\sigma _4 \sigma _3 *} 
c_{j \sigma _3 } c_{i \sigma_4 }
+
h.c. 
)-\mu\sum_{i  }c_{i \sigma } ^\dagger c_{i \sigma },
\end{alignat*} 
where the order parameters 
($
\Delta_{ij}^{\sigma \sigma '\,}
= 
-\Delta_{ji}^{\sigma '\, \sigma }
$)
are given by 
$
{\Delta} _{ij}^{\sigma _1 \sigma _2 } = 
V_{ij}^{\sigma_1 \sigma_2; \sigma_3 \sigma_4 } 
\langle 
 c_{j \sigma _3} c _{i\sigma_4}
\rangle 
$.
The usual mean field theory leads to the gap equation of which  
a solution gives an order parameter. 
Here we do not follow this procedure but
 a priori assume order parameters
which may be realized for some  interactions 
$V_{ij}^{\sigma_1 \sigma_2 \sigma_3 \sigma_4 }$.
Let us consider  the two cases separately\cite{vol-g,ueda-sg}:
(i) singlet states
$
\mb{ \Delta}_{ij} =-\tilde{ \mb{ \Delta}}_{ij} 
=  \psi_{ij}i \sigma _y
$, 
$
(\psi_{ij} = \psi_{ji})
$ 
 and (ii) triplet states
$
\mb{ \Delta}_{ij} =\tilde{\mb{ \Delta}}_{ij} 
=  ( \mb{d}  _{ij}\cdot \mb{\sigma} ) i \sigma _y
$,
$
(\mb{d} _{ij} =  - \mb{d}  _{ji})
$, 
where
$ (\mb{\Delta}_{ij})^{\sigma\sigma'}=\Delta_{ij}^{\sigma\sigma'}$
is a $2\times 2$ matrix in the spin space
 and
$\tilde {\ }$  denotes matrix transpose.
Now  assume the translational symmetry, namely,
$ t_{ij} = t(i-j)$,
$\mb{\Delta}_{ij} =\mb{\Delta}(i-j)$ and also
the absence of a magnetic field,  that is, $t(i-j)$ to be real.
Then,  
except a constant, the BdG hamiltonian is given by 
a $4\times 4$ matrix $\mb{h}_k $ as 
\begin{alignat*}{1} 
{\cal H} 
=&  \sum_k \mb{c} ^\dagger _{{k}} \mb{h} _{{k}} \mb{c}_{{k}},
\quad
\mb{h} _{{k}}
= 
\mat 
{\epsilon_{{ k}} \sigma _0   } { \mb{ \Delta} _{{k}}}
{\mb{\Delta }^\dagger_{{k}}}{-\epsilon_{{k}}  \sigma _0 } 
\end{alignat*}  
where 
$
\mb{c} ^\dagger _{{k}}  = 
(c ^\dagger _{\uparrow }(\mb{k} ),
c ^\dagger _{   \downarrow }(\mb{k} ),
c  _{ \uparrow }(-\mb{k} ),
c  _{  \downarrow }(-\mb{k} ))
$
with 
$c_\sigma (\mb{k} ) = \frac {1}{\sqrt{V}} \sum_j e^{i \mb{k} \cdot \mb{r} _j} c_{j \sigma }$,
$
\epsilon_{{k}}=\sum_{\ell} e^{  -i  \mb{k}    \cdot  \mb{r} _\ell}
t(\ell)-\mu
$,
$
\mb{\Delta}_{{k}}
=\sum_{\ell} 
e^{
- i \mb{k}    \cdot \mb{ r}_\ell  
}
\mb{ \Delta} (\ell)
$,
$
\mb{ \Delta } _{-{k} } =- \tilde {\mb{ \Delta }}_{{k} }
$, 
$\sigma _0=\mat{1}{0}{0}{1}  $, and $V$ is a volume of the sysmtem.
The order parameter is given by
$
\mb{\Delta}_{{k} } 
=   \psi_ k i \sigma _y
$,
$
\tilde {\mb{ \Delta }}_{{k}}  = -{\mb{ \Delta }} _k
$ for singlet states  and 
$
\mb{\Delta}_{{k} }  
=  ( \mb{d} _{{k}}\cdot \mb{\sigma }) i \sigma _y
$,
$
\tilde {\mb{ \Delta }}_k = \mb{ \Delta }_k 
$ for triplet states
( $\psi_{ {k}} $ is even and 
$ \mb{d}_{{k}}$ is odd in $\mb{k} $).
\footnote{ 
The sum 
is over a half of the 
Brillouin zone to avoid double counting.
}

The BdG hamiltonian
has a  particle-hole symmetry.
If 
$
\mb{h}_k \mvec{\mb{u}_k }{\mb{v}_k }
=E_k\mvec{\mb{u}_k }{\mb{v}_k }
$,
then $C\mvec{\mb{u}_k }{\mb{v}_k }$ 
is also an eigenstate with energy $-E_k$ where
$C=\rho_x K$ for singlet states and 
$C=-i\rho_y K$ for triplet states
 ( $\mb{u}_k$ and $ \mb{v}_k  $ are the  two-component vectors and 
$K$ is a complex conjugate operator and 
the Pauli matrices 
$\mb{\rho}$ operate on the  two component blocks.)
 \footnote{
The $4\times 4$ matrices are
 spanned by $\rho_i \otimes \sigma_j,i,j=0,\cdots, 3$.
} 
Then  it is useful to consider 
$\mb{h}_k^2=
\epsilon_{k}^2\rho_0+
 \mat{ \mb{ \Delta}_{k} \mb{\Delta}_{k} ^\dagger}{0}{0}
{ \mb{ \Delta}_{k} ^\dagger \mb{\Delta}_{k}   }$.
For singlet states, we have
$ \mb{ \Delta}_{k} \mb{\Delta}_{k} ^\dagger 
 =  \mb{ \Delta}_{k} ^\dagger \mb{\Delta}_{k}
= |\psi_k|^2\sigma_0
$
and for triplet states,
$ \mb{ \Delta}_{k} \mb{\Delta}_{k} ^\dagger  =
|\mb{d}_k|^2\sigma_0 + \mb{q}_k\cdot \mb{\sigma } $
 with a real vector
$\mb{q}_k=i \mb{d}_k \times \mb{d}_k^*  $,
which we call $q$-helicity ( $\dagger$ represents  hermite conjugate and $^*$ 
 complex conjugate).

{\em Chern numbers for unitary  states}.
Singlet order and triplet order with vanishing $q$-helicity 
are called unitary since 
 $ \mb{ \Delta}_{k} \mb{\Delta}_{k} ^\dagger  =
\mb{\Delta}_{k} ^\dagger   \mb{ \Delta}_{k} \propto \sigma _0 $.
Nonunitary triplet states ($\mb{q}_k\neq 0 $) will be discussed later.
For unitary states, we define a  unitary matrix
$\mb{\Delta }^0_k $ by 
$\mb{\Delta}_k = |\Delta_k | \mb{\Delta}^{0}_k$, 
($|\Delta_k| =|\psi_k|$ for singlet states and
$|\Delta_k|= | \mb{d}_k |$ for triplet states, respectively).
Since the spectra are doubly degenerate as will be shown later,
 fixing phases of the states
is not enough to determine Chern numbers by the standard procedure
\cite{mkann,yh-qhe-review,chern_yh}.
Instead, one can define non-Abelian vector potentials and fluxes 
following definitions of generalized non-Abelian connections
\cite{zee-wl}.

Let us assume that the states are $M$-fold degenerate ( $M=2$ in the present 
unitary case)
as
$|  \alpha \rangle$,
$\alpha = 1,\cdots, M
$.
Then a non-Abelian connection is defined by  
$A^{\alpha\beta }_\mu = 
 \langle \alpha   |\partial _\mu| \beta  \rangle
$,
$
{\cal A}^{\alpha \beta } = A^{\alpha \beta }_\mu dk_\mu
$ and 
$\partial _\mu=\partial_{ k_\mu},\mu=x,y,z
$.\footnote{ 
Summation over the repeated indices $\mu $
 is assumed. 
}
A unitary transformation of a
degenerate state
 $| \alpha  \rangle \to  | \bar \alpha \rangle 
 =   | \alpha  \rangle \omega^{\alpha \bar \alpha } 
$
($\mb{\omega}$: unitary)
causes  ``a gauge transformation''
$
\mb{\cal\bar A}  = 
\mb{\omega}^\dagger  \mb{\cal A} \mb{\omega} + \mb{\omega} ^\dagger d \mb{\omega} 
$.
Then the field strength
$\mb{\cal F}  = d \mb{\cal A} + \mb{\cal A} \wedge \mb{\cal A}
$
is gauge covariant since
$
\mb{\cal \bar F}  
= \mb{\omega} ^\dagger  \mb{\cal F} \mb{\omega} 
$.
One may write
$\mb{\cal F} 
=  \frac {1}{2!} \mb{F} _{\mu\nu} dk_\mu \wedge dk_\nu 
$,
$
\mb{F} _{\mu\nu} = \partial _\mu \mb{A} _\nu - \partial _\nu \mb{A} _\mu
+ [ \mb{A} _\mu,  \mb{A} _\nu]
$, ($(\mb{A}_\mu)^{\alpha \beta } = A_\mu^{\alpha \beta }$).
Then ``a magnetic field'' in the $\mu$-direction 
 is $
{B}_\mu 
=  \frac {1}{2} \epsilon_{\mu\nu\lambda} {\rm Tr \,} \mb{F}_{\nu\lambda} 
$.
Since ${\rm Tr \,} \mb{\cal F} $ is  unitary invariant,
so is ${B}_\mu $.
The total flux passing through the $\nu\lambda$-plane
is given by an  integral
of the magnetic field  ${B}_\mu $ 
 over the two-dimensional 
 Brillouin zone, $T^2_{\nu\lambda}$,
( $k_\mu$ is fixed).
The first Chern number
 is \cite{eguchi}
\footnote{ 
$T^2_{\nu\lambda}=\{(k_\nu,k_\lambda)\  |k_\nu,k_\lambda\in [0,2\pi)\} $ and 
$k_\mu$ runs over $[0,\pi]$ to avoid double counting.
},
\begin{alignat*}{1} 
C_{\mu}(k_\mu) =&
 \frac {1}{2!} \epsilon _{\mu\nu\lambda}
\frac {1}{2\pi i}  \int_{T^2_{\nu\lambda}} {\rm Tr \,} \mb{\cal F} 
=
 \frac {1}{2!} 
\frac {1}{2\pi i}  \int_{T^2_{\nu\lambda}}dk_\nu\wedge dk_\lambda\,  { B} _\mu. 
\end{alignat*} 
This is the Abelian chiral anomaly discussed in the 
non-Abelian gauge theories\cite{eguchi,zumino}.
Here we have considered
 the  cubic lattice. 
Extensions to other lattice structures is
straightforward.

{\em Dirac monopole in the parameter space}.
The BdG equation for the unitary states 
$\mat
{{\epsilon}_k\sigma_0  }{{|\Delta}_k| \mb{\Delta}_k^{0}  }
{{|\Delta}_k| (\mb{\Delta}_k^{0}) ^{-1}  } { - {\epsilon}_k\sigma_0  }
\mvec
{\mb{u}_k} 
{\mb{v}_k} 
= 
E_k
\mvec
{\mb{u}_k} 
{\mb{v}_k}$
reduces to  
\begin{alignat*}{1} 
\mat
{{\epsilon_k} }{|\Delta_k| }
 {|\Delta_k | } { - {\epsilon}_k } _\rho
\otimes \sigma_0 
\mvec
{\mb{u}_k} 
{ \mb{\Delta} _k^0\mb{v}_k} 
=& 
E_k
\mvec
{\mb{u}_k} 
{ \mb{\Delta}_k^0\mb{v}_k}.  
\end{alignat*}
Thus  energies 
are given by 
$
 E_k = \pm R
$
($
R = \sqrt{\epsilon_k ^2+ |{\Delta}_k |^2}
$)
and the states are doubly degenerate.
The band with energy $-R$ is the superconducting condensates of pairs. 
On the other hand, the band with energy $+R$ represents quasiparticle excitations.
By  a parametrization:
$
\epsilon_k = R \cos\theta
$ and
$
|\mb{\Delta}_k | = R \sin\theta$,
eigenvectors of condensate ($E_k=-R$) are 
$\mvec
{ \mb{u}_k  }
{\mb{\Delta} _k^0 \mb{v} _k }
=
{| R,\theta \rangle }_\rho
\otimes
{| \alpha  \rangle }_\sigma$,
where 
$| R,\theta\rangle_\rho = 
\mvec
{-\sin \frac \theta 2}
{\cos \frac \theta 2}
$
and $| \alpha   \rangle_\sigma  $,
 ($\alpha =1,2$)
are arbitrary orthonormalized states in the $\sigma $-space. 
Now let us take eigenstates 
 of the 
$\mb{\Delta }_k^0$ with eigenvalues $e ^{-i\phi_\alpha}$ ($\alpha =1,2$)
 to calculate the Chern numbers.
\footnote{ 
It is a similar procedure to fix  phases of states
 for non-degenerate cases
\cite{mkann}.
}
The degenerate orthonormal eigenvetor
for the condensate band $E_k=-R$ is given by
$
|\psi_\alpha \rangle =
\mvec
{\mb{u}_k }
{\mb{v}_k }_ \alpha 
= 
\mvec
{-\sin\frac \theta 2 | \alpha  \rangle_\sigma  }
{e^{i\phi_\alpha } \cos\frac \theta 2 | \alpha  \rangle_\sigma  }
$\cite{yh-sr}.
The connection is given by
\begin{alignat*}{1} 
A_\mu^{ \alpha \beta } =& 
\langle \psi_\alpha  |  \partial _\mu | \psi_\beta  \rangle 
= 
\langle  \mb{R}  _{\alpha} | \partial _\mu | \mb{R}  _{\beta}  \rangle_\rho
 \langle   \alpha | \beta \rangle_\sigma  
+ 
\langle  \mb{R}  _{\alpha} |  \mb{R} _{\beta}  \rangle_\rho
\langle \alpha | \partial _\mu | \beta \rangle _\sigma ,
\\
=&  
A_{\mu}^{ \alpha \beta  } (\rho) \delta_{\alpha \beta } 
+ 
\langle  \mb{R}  _{\alpha} |  \mb{R}  _{\beta}  \rangle_\rho
A_{\mu}^{ \alpha \beta   } (\sigma ),
\end{alignat*} 
where
$
A_{\mu}^{ \alpha \beta } (\rho)
= 
\langle  \mb{R}  _{\alpha} | \partial _\mu |  \mb{R}  _{\beta}  \rangle_\rho
$ with
$
| \mb{R} _{\alpha } \rangle_\rho = | R,\theta,\phi_\alpha  \rangle_\rho 
=\mvec
{-\sin\frac \theta 2 }
{e^{i\phi_\alpha } \cos\frac \theta 2 }
$.
Also 
$
A_\mu^{\alpha \beta }( \sigma ) = 
\langle \alpha | \partial _\mu | \beta \rangle _\sigma 
= ( \mb{U}^\dagger  \partial_\mu \mb{U} )_{\alpha \beta }
$
with
$
\mb{U}  = (|1\rangle_\sigma  , |2 \rangle_\sigma  ) 
$.
Then 
the total magnetic field in the parameter space 
is 
$
{B}_{\mu} = 
{B}_{\mu} (\rho) 
+ {B}_{\mu} (\sigma ) 
$,
where ${B}_{\mu} (\rho) = \epsilon_{\mu\nu\lambda} {\rm Tr \,} 
 \partial_\nu \mb{A}_\lambda(\rho)$
 and $ {B}_{\mu} (\sigma ) 
=
\epsilon_{\mu\nu\lambda} {\rm Tr \,} 
 \partial_\nu \mb{A}_\lambda(\sigma )$. 
Since $B_\mu(\sigma )$ vanishes by the ``sum rule'' among the filled bands,
we have
$
{B}_{\mu} 
= 
{B}_{\mu} (\rho)$.
\footnote{ 
$
B_\mu(\sigma )
=\epsilon _{\mu\nu\lambda}{\rm Tr }
\partial _\nu( \mb{U} ^{-1}  \partial_\lambda \mb{U} )
=   
- \epsilon _{\mu\nu\lambda}{\rm Tr }
( \mb{U} ^{-1} \partial _\nu \mb{U} )(\mb{U} ^{-1}\partial_\lambda \mb{U})
= 0$.
It also completes a general proof of the sum rule 
for the non degenerate states,
that is,  the 
sum of the Chern numbers for all bands is zero.
}
It implies that
the Chern numbers, $C_{\alpha ,\mu}$, 
 of the condensed band   in the $\mu$-direction,
is given by the sum of the 
Chern numbers of the two 
vectors
$
| R _\alpha \rangle _\rho
$,
$(\alpha =1,2)$ 
which are the eigenstates
of the 
$2\times 2$ hamiltonians
\begin{alignat*}{1} 
\mb{h}_k^\alpha =& 
 \mat
{\epsilon_k }{e^{i\phi_\alpha }|\mb{\Delta}_k |}
{e^{-i\phi_\alpha }|\mb{\Delta}_k |}{- \epsilon_k }
=  \mb{\sigma}  \cdot \mb{R}_\alpha,
\end{alignat*} 
where $ R_{\alpha, X} = R \sin\theta\cos\phi_\alpha$,
$
R_{\alpha, Y} = R \sin\theta\sin\phi_\alpha  
$ 
and 
$
R_{\alpha, Z} = R \cos\theta
$.
Namely they are
$C_\mu = \sum_\alpha C_{\alpha ,\mu}$.
Now we have 
reduced the problem 
to calculate the Chern numbers of the
eigenstates of the $2\times 2$ matrices, $\mb{h}_k^\alpha  $.
By mapping from the two-dimensional Brillouin zone to
the three-dimensional space,
$T^2_{\nu\lambda} \ni (k_\nu,k_\lambda)\to \mb{R}_\alpha   $,
we obtain a closed oriented surface $R_\alpha ( T^2_{\nu\lambda})$. 
The wrapping degree of the map around the origin 
gives a charge of the Dirac monopole sitting there.
This is the Chern number $C_{\mu}(k_\mu)$\cite{yh-sr,yh-sr-lt}.

For the present degenerate case,
the map from a two-dimensional point to 
{\em two} three-dimensional points
$
T^2_{\nu\lambda} \ni (k_\nu,k_\lambda)\to \{\mb{R}_{\alpha=1},\mb{R}_{\alpha=2}\}
$
defines ( fixing $k_\mu$) the
 surfaces
$
\{\mb{R}_{\alpha=1}(T^2_{\nu\lambda}),\mb{R}_{\alpha=2}(T^2_{\nu\lambda})\}
$
which determine the two covering degrees of the maps around the origins,
$
 N_{\alpha ;\nu\lambda}(k_\mu)
$,
$
(\alpha =1,2)
$.
They 
give 
the Chern numbers
$C_{\alpha,\mu} $ respectively. 
Since only the
condensed states
are filled 
for the superconducting ground state, 
the Chern numbers
of  the unitary states are given
by
\begin{alignat*}{1} 
C_\mu(k_\mu)  =& \frac {1}{2!}\epsilon_{\mu\nu\lambda}  N_{\nu\lambda}(k_\mu),
\quad
 N_{\nu\lambda}(k_\mu) = 
\sum_\alpha  N_{\alpha ;\nu\lambda}(k_\mu).
\end{alignat*} 
The Chern numbers defined here for the 
unitary superconductors satisfy
 $C_{\nu\lambda}(k_\mu)=4\times {\rm  integer} $
 for the singlet order
and $ C_{\nu\lambda} (k_\mu)=2\times{\rm  integer} $ for the triplet order.
\footnote{  
For the singlet case, 
we have $N_{1;\nu\lambda}=N_{2;\nu\lambda}$,
since $\phi_1=\phi_2+\pi$.
Further  each $N_{2;\nu\lambda}$ is even\cite{yh-sr,yh-sr-lt}. 
For the triplet case, $N_{1;\nu\lambda} \neq N_{2;\nu\lambda}$ generically.
However, the some of them is even since we can show
 $\phi_1=+\vartheta+ {\rm Arg } \delta $,
 $\phi_2=-\vartheta+ {\rm Arg } \delta +\pi $ with some function $\vartheta$
where $\mb{d}\cdot \mb{\sigma} $ is unitary equivalent to
$ \delta \sigma_z$ $(| \delta | = |\mb{d} |) $.
We use a base which diagonalizes $\mb{\Delta}_{k}^0 $. 
Another base which diagonalizes
$\mb{d}\cdot \mb{\sigma } $ is also useful. 
We thank A. Vishwanath for the communication on this point.
}

{\em Nonunitary states}. 
In these triplet states,
there are no degeneracies in solutions of 
 the BdG equation.
There are  four quasiparticle bands,
which are classified by the
$q$-helicity as 
\begin{alignat*}{1} 
& (\mb{\sigma} \cdot \mb{ q}_k ) \mb{u}_\pm 
= \pm q_k  \mb{u}_\pm,\quad q_k=|\mb{q}_k |,
\\
 {\mb{h}}^2_k
\psi_j^\pm
&= ( \epsilon_k ^2+ | \mb{d}_k |^2 \pm q_k) 
\psi_j^\pm,\\ 
\psi_1^\pm =& \mvec
{\mb{u} _\pm}
{\mb{v} _\pm},
\psi_2^\pm=\mvec
{\mb{u} _\pm}
{-\mb{v} _\pm},\quad 
\mb{v}_\pm  =  -i  \sigma_y \mb{u} _\mp.
\end{alignat*} 
Then  states with helicity $+q_k$ and energy 
$\pm E_{+q} = \pm \sqrt{\epsilon_{k} ^2+|\mb{d}_k |^2 +q_k}$,
are
$
|\pm  E_{+q} \rangle 
=  \mb{U} _+ \mb{\eta}^\pm_{+q}
$, where
$
\mb{U} _+= 
 \frac {1}{\sqrt{2} }
\mat
{\mb{u}_+}{\mb{u}_+}
{\mb{v}_+}{- \mb{v}_+}
$
and 
the orthonormal vectors $\eta^\pm_{+q}$ are determined as 
the  eigenvectors of the reduced $2\times 2$ hamiltonian 
$
\hat {\mb{h}}_{+q} = \mb{ U}_{+q} ^\dagger  {\mb{h}} \mb{U} _{+q}
$
with energies $\pm E_{+q}$.
The hamiltonian $\hat {\mb{h}} _{+q}$ is traceless 
and it can be  expressed  by
the Pauli matrices, 
$
\hat {\mb{h} } _{+q} =  
\mb{ \sigma} \cdot \mb{  R}_{+q}
$
where $\mb{R}_{+q} =
( \epsilon ,  {\rm Im\,}  d _{+-},
 {\rm Re\,} d _{+-}) $ is a real vector and 
$
d _{+-} =
 \mb{u} _+ ^\dagger( \mb{d}_k \cdot \mb{\sigma })  \mb{u} _-
$.
As for the helicity $-q_k$ state, 
one can follow 
almost the same procedure and obtain
the reduced BdG hamiltonian similarly.
Here 
we can define several topological invariants.
As discussed above,
 the states with $\pm q_k$-helicities and energies 
$\pm E_{\pm q}$ are non-degenerate,
\begin{alignat*}{1} 
\mb{h}_{ k } |     \varepsilon_{e}  E_{\varepsilon_{q}} \rangle 
=&   \varepsilon_{e}  E_{\varepsilon _q } |  \epsilon_{e}  E_{\varepsilon_{q}} \rangle, 
\mb{Q}_k 
  |   \varepsilon_{e}  E_{\varepsilon_{q}} \rangle
=  \varepsilon_{q}  |    \varepsilon_{e}  E_{\epsilon_{q}} \rangle,
\\
\mb{Q}_k =&  
\mat  
{ \mb{\sigma }\cdot \mb{q}_k}{O}
{O}{-\sigma _y\mb{\sigma }\cdot \mb{q}_k  \sigma _y},
[\mb{h}_k, \mb{Q} _k ]=  0,
\end{alignat*} 
where 
$E_{\varepsilon_{q}q}  ( \mb{k} ) = \sqrt{|\epsilon _{k}|^2 + 
| \mb{d}  _k |^2 +\varepsilon_{q}  }
$,
($ \varepsilon_{q} =\pm q_k$ and $ \varepsilon_{e} =\pm $).
Then the standard Chern number $C_\mu^0(k_\mu)$
 in the $\mu$-direction for a fixed $k_\mu$ 
is obtained by the standard way\cite{mkann}.

Also we have topological invariants in the $\mu$-direction 
$N^q_{\mu}(k_\mu)$.
They are 
 wrapping degrees around the origin of the 
 map
$
(k_\nu,k_\lambda)  \to \mb{q}(\mb{k} ) = \mb{q}_k 
$,
which  define  closed surfaces 
$\mb{q} (T^2_{\nu\lambda})$ in three-dimensions\cite{yh-sr}.
Further we have other topological invariants
$ N^\pm_{\mu}(k_\mu)$,
 which are also  wrapping degrees of the map around the origin,
$(k_\nu,k_\lambda)  \to \mb{R}_{\pm q} (\mb{k} )$.
(The reduced hamiltonians are
$\tilde{ \mb{h}} _{\pm q} = \mb{R}_{\pm q} \cdot \mb{\sigma }  $ 
for the $q$-helicity $\pm q_k$.)

Up to this point, 
topological arguments have been applied to gapful cases.
However,
the nodal structure of the gap function is, in fact,
characterized by the topological description.
Formally we have treated a three-dimensional superconductivity as 
a collection of two-dimensional systems
parametrized by, say, $k_z$.
In $ \mb{R}$-space, closed surfaces parametrized by 
$(k_x,k_y)$ are generically away from the  monopole at the origin. 
As $k_z$ is varied,
the surfaces move around and they can pass through the
monopole.
 Since the distance between a
point on a surface and the monopole gives 
a half of the energy gap $E_g(k_x,k_y;k_z)$,
the gap closes at the value of $k_z$ when the monopole is on the surface.
Thus the  nodal structure of the superconductivity
is {\em point-like  generically} if the time-reversal symmetry is broken. 
When the Chern number
jumps as $k_z$ varies, 
the superconducting gap has to be closed due to  
topological stability of
Chern numbers.
Also a non-zero Chern number implies that 
 the corresponding two-dimensional system has 
a non-trivial topological order. 
The superconducting node 
is considered as the critical point of the quantum phase transition
between two states with different toopological orders.
\begin{figure}
\includegraphics[width=8.6cm]{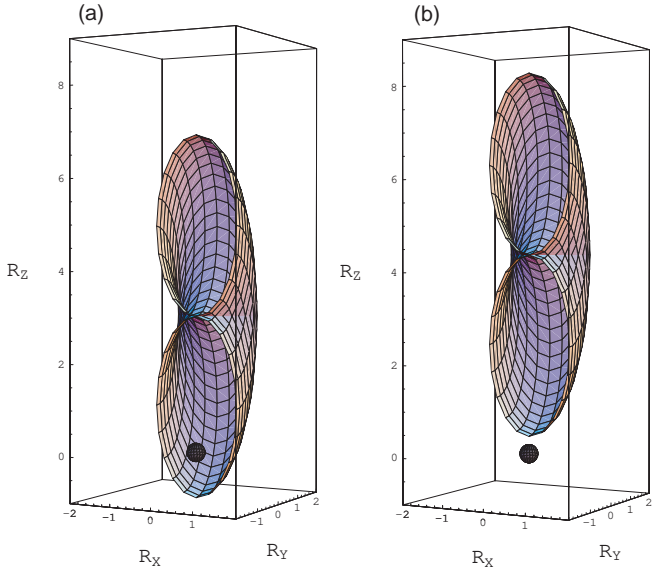}%
 \caption{
Examples of closed surfaces 
$R_1(T^2_{xy})$ which are cut by the $R_Y-R_Z$-plane.
The  monopole is at the origin.
$t=d_{z}= 1$, $\mu=-5$: (a)$k_z=0$, (b)$k_z=-2 \pi/5$.
\label{fig1}}
 \end{figure}

 To make the discussion clear, let us take an example 
$
\mb{ \Delta } _{{k}} = d_z^0( \sin k_x+ i \sin k_y) \sigma _x 
$,
$
\epsilon_{{k} }   = -2 t ( \cos k_z+\cos k_y+ \cos k_z)-\mu$, ($ t>0$)
 \cite{yh-sr}.
This is an analogue of the
Anderson-Brinkman-Morel(ABM) state in $^3$He superfluid.
For a fixed value of $k_z$, the surface is reduced to that of
the chiral $p$-wave order parameter with a modified chemical potential,
$\mu-2t \cos k_z$.
( we can recover the ABM state, $\mb{d}_k \to (0,0,d_z^0(k_x+ik_y)) $
 in the limit of $\mu\to-6t+0, \mb{k} \to 0  $.)
There are two quantum phase transitions
which are accompanied by jumps of the Chern number between 
$-2$ and $0$ for $-6t < \mu<-2t$.
These correspond to the gap nodes at the north and south poles on the
Fermi surface.
In Fig. 1,  surfaces $ \mb{R}_1(T_{xy }^2) $ 
are shown with the monopole at the origin.
The Chern number jumps between  $-2$ and $+2$ for $-2t < \mu<2t$ and 
$0$ and $+2$ for $2t < \mu<6t$.

For line nodes, we need  
{\em additional
constraints }
to keep the  monopole on the closed surfaces
when $k_z$ is varied.
To make a discussion simple, we take singlet order 
or  triplet order with  $d_x=d_y=0$ and $d_z \neq 0$.
Further let us require that  the order parameters are 
real,
 namely
 we have a {\em time-reversal symmetry}.
Then the closed surfaces in the $\mb{R}$-space collapse 
into a board like region 
on the $R_X$-$R_Z$ plane and
one can expect a situation where the  monopole
moves along the surface when $k_z$ is varied.\cite{yh-sr-lt} 
Thus a line node appears in the superconducting gap. 
As shown in this example, the  nodal structure of the superconductivity
has a fundamental relation to topological order.
The detailed discussion 
on this point will be given elsewhere.\cite{yh-mk}


We had useful discussions with M. Sato.
Part of the  work by Y.H. was supported by Grant-in-Aid from
Japanese Ministry of Science and Culture and 
the KAWASAKI STEEL 21st Century Foundation. 
 The work by S.R.  was in part supported by JSPS.



\end{document}